# Fabrication and characterization of shape- and topology-optimized optical cavities with deep sub-wavelength confinement for interfacing with colloidal quantum dots

Mohammad Abutoama,[1,2,*,#] Rasmus Ellebæk Christiansen,[2,3,**,#], Adrian Holm Dubré[1,2], Meng Xiong,[1,2] Jesper Mørk,[1,2] and Philip Trøst Kristensen[1,2]

[1]Department of Electrical and Photonics Engineering, Technical University of Denmark, Ørsteds Plads, building 343, 2800 Kgs. Lyngby, Denmark
[2]NanoPhoton - Center for Nanonphotonics, Ørsteds Plads, building 345A, 2800 Kgs. Lyngby, Denmark
[3]Department of Civil and Mechanical Engineering, Technical University of Denmark, Nils Koppels Allé, Building 404, 2800 Kgs. Lyngby, Denmark
*moabo@fotonik.dtu.dk
**raelch@dtu.dk
#Equal contribution

## Abstract

We employ a combined shape- and topology-optimization strategy to design manufacturable two-dimensional photonic crystal-based optical nanocavities that confine light to length scales well below the resonance wavelength. We present details of the design strategy as well as scanning electron micrographs of the fabricated indium phosphide cavities with a compact footprint of ~4.5λ×4.5λ, which feature gaps on the order of 10 nm and theoretical mode volumes in the gap center below 0.1 $(\lambda/2n_{air})^3$. Subsequent optical characterization of the far-field emission as well as Purcell-enhanced photoluminescence from the cavities with and without spin-coated colloidal quantum dots are compared to numerical simulations. The results corroborate the potential of the design strategy and fabrication process for ensuring high yield and reliable performance as well as the viability of the material platform for exploring light-matter interaction with colloidal QDs.

## Introduction

Two main mechanisms can be used to enhance the light-matter interaction in nanostructures; the first one is by confining the light spectrally, increasing the quality factor Q of the resonator, while the second one is by confining it spatially, shrinking the resonator mode volume V [1]. Q-values up to several million [2] were achieved with two-dimensional photonic crystal (2D PhC) dielectric cavities. This significant improvement was achieved by breaking symmetry or reducing structural disorder [3-4] through a shifting of the air holes in the central part or changing their sizes [3, 5], or by forming a nanocavity with a locally modulated waveguide [6]. Many of the works on 2D PhC structures were based on variations of so-called $L_p$ line defect-cavities [3] where p is the number of removed inclusions in the PhC. In all these examples, Q was improved by several orders of magnitude, but with modest spatial confinement typically larger than the so-called diffraction limit $V\sim(\lambda/2n)^3$ [7-8], where λ is the resonance wavelength and $n$ is the refractive index. To enhance the light-matter interaction in the Purcell enhancement [9] regime while maintaining a large bandwidth, it can be interesting to consider the possibilities of optical nanocavities with modest Q and very small V. Generally, this is challenging because a small V requires extremely small features in the cavity center. While it is well known that plasmonic nanostructures can confine light in the deep subwavelength regime, the quality factors are significantly lower than the ones in dielectrics [10] due to the unavoidable losses in metals [7].

As an alternative to plasmonic resonators, there has lately been a considerable interest in dielectric bowtie cavities with deep subwavelength confinement [11-20]. The bowtie geometry exploits the discontinuous nature of the electric field as dictated by the boundary conditions in the macroscopic Maxwell equations [15-17] to locally enhance the relative electric field strength. The resulting orders-of-magnitude reduction in mode volume enables strong light-matter interaction without resorting to high quality factors. This is important for reducing the footprint and energy requirements and for applications requiring relatively large bandwidths such as nano light-emitting diodes with noise squeezing for optical interconnects applications [21-22] and quantum optics with broadband emitters for quantum optics [23]. An important property of an optical cavity, to be specified prior to design, is whether the light is to be confined in air or inside the material with different applications requiring different confinement. Different examples include confinement inside the material for interaction with emitters, such as embedded quantum dots [24] or color centers [25] and with applications as efficient single photon sources or for cavity quantum electrodynamics. Similarly, strong confinement in air is useful for interfacing with extrinsic emitters, such as colloidal quantum dots [26] or molecules [27], or in connection with subsequent deposition of optically active materials for non-linear signal processing [16] or sensing [28].

While many cavity designs rely on physical or mathematical insights, a completely different avenue is offered through inverse design by topology optimization [29-32] to maximize the local density of states (LDOS). Interestingly, the inverse design procedure in these structures naturally results in dielectric bowtie cavities. In some of the first theoretical studies in the field [11-12], fabricability was disregarded, providing interesting proof-of-concepts but leading to unrealistic device blueprints. Once the limitations of materials and fabrication are considered, realistic designs can then be obtained [31, 33], and it was shown in Ref. [31] that for small topology-optimized cavities with realistic fabrication constraints, the ratio Q over V could be enhanced by up to two orders of magnitude relative to standard $L_1$ cavities and more than one order of magnitude relative to shape-optimized $L_1$ cavities. The strong interplay of design and fabrication in this field was recently shown in Ref. [34], where manufacturing constraints of a state-of-the-art process were included in the topology optimization algorithm to ensure successful fabrication of the resulting design with deep subwavelength confinement inside the material in the cavity center. For limited footprints, moreover, there appears to be an interesting tradeoff between Q and V for achieving high Purcell factors. Indeed, it was shown in Ref. [35] that the highest Q/V -ratio does not necessarily have the highest Q or the lowest V. The effect of the cavity size is seen through the fact that for large footprint, the enhancement of the Q/V-ratio is achieved by the geometry supporting a larger Q at the cost of a larger V, while for the small footprint cavities, similar Q/V-ratio can be obtained by supporting a significantly smaller V at the cost of a lower Q.

For additional context, we review a few approaches for localizing light to an air region in dielectric 2D PhCs. In Ref. [36] for example, V~ 0.1 $(\lambda/2n_{air})^3$ and Q ~$10^4$ was achieved by creating a defect through reducing the radius of the central hole in a hexagonal cavity. In Ref. [37], Fourier space analysis was used to analyze the symmetry of the defects and a graded square lattice with V~ 0.3 $(\lambda/2n_{air})^3$ and Q ~$10^5$ was obtained; an experimental demonstration was subsequently shown in Ref. [38] with Q~$1.3 \cdot 10^4$. The first dielectric bowtie cavity with confinement in air was proposed in Ref. [11] using inverse design. Starting from completely random patterns, the result was a spontaneous emergence of periodical patterns but with unrealistic designs for fabrication. The design has V~ 0.112 $(\lambda/2n_{air})^3$ and Q~300. In Ref. [39], a gap mode was demonstrated experimentally in a non-terminated air-slot PhC with Q~$10^4$ and V~0.16 $(\lambda/2n_{air})^3$. This was proposed based on the original work in Ref. [40] where an air-slot was formed by shifting the air holes away from the waveguide. The slot introduces a dielectric discontinuity, which helps to enhance the field at the air region. An air-slot type cavity was implemented in Refs. [41-42] for optomechanical coupling applications. Usually, in the air-slot and air-hole defect designs, the footprint of the cavity size was not considered in the design process and consequently large-footprint cavities were proposed in many of the works. An $L_1$ cavity was topology-optimized in Ref. [31] with length-scale control to systematically propose cavities for Purcell enhancement applications with significantly reduced V~0.0021 $(\lambda/2n_{air})^3$ and Q~1062. These device metrics relied on few-nanometer sized features in the cavity making it impossible to realize using state of the art fabrication tools. We note, that in addition to 2D PhC-based cavities – such as the ones we focused on in this work – optical cavities based on nanobeams has been shown to provide remarkable possibilities [15, 17]. For the present study, however, it is not clear that they are compatible with the spin-coating of colloidal quantum dots (QDs).

Colloidal QDs are semiconductor nanocrystals which are chemically grown in solution to have a typical diameter ranging from a few nanometers up to roughly 10 nm. They have received significant interest in various fields due to their interesting properties and applications [43-45]. Among their advantages are high quantum yield, the tunability of the emission wavelength by changing the QD size and composition, a reduced re-absorption process due to the existence of the Stokes shift between the excitation and emission wavelengths [44], and their ability to act as single photon sources at room temperature [44]. One of the major obstacles for using them in practical applications is the requirement for accurate positioning of a single QD in the hotspot of the cavity to achieve an efficient interaction with the light [26, 44-47].

Here, we detail the design, modeling, fabrication, and characterization of a geometrically simple cavity with deep subwavelength confinement tailored for confining the light in air with the purpose of interfacing the resonant optical field with colloidal quantum dots [43-47]. Hence, we shape- and topology-optimized a manufactural 2D PhC to maximize the Purcell-enhancement at the center of a cavity with footprints on the order of the wavelength. We characterized the samples using photoluminescence (PL) with and without QDs as a proof of principle for the viability of the material platform for interfacing colloidal QDs with this type of optical cavity with subwavelength confinement.

## Design, fabrication, and modeling

We utilized the sensitivity of the resonator mode volumes to sub-wavelength geometric details near the optical hotspot [48] while enhancing the mode quality factor by optimizing larger, easily fabricated features far from the hotspot. Specifically, we explored the application of a hybrid optimization approach to design the 2D layout of a membranized device. This eases fabrication and simultaneously enables the inverse design and optimization of significantly larger devices than is computationally tractable using traditional density-based topology-optimization, which most often rely on an extremely fine discretization of the design to ensure sufficient geometric freedom. This, in turn, poses a computational barrier to considering larger device sizes. We effectively circumvented this problem by employing shape-optimization of a predefined topology for most of the design domain. We restricted the density-based topology-optimization to the central cavity region which constitutes approximately 5% of the overall device footprint, where we were able to employ a minimum in-plane design resolution of 5 nm x 5 nm and as small as 1 nm x 1 nm close to the center, while employing shape-optimization for the fixed topology given by an initial hexagonal 2D PhC pattern in regions further away. Our starting point was the unoptimized 2D PhC cavity with a footprint of $\sim 4.5\lambda \times 4.5\lambda$ shown in Fig. 1(a) which supports a cavity mode with $V_{r_0} \approx 1.92\ (\lambda/(2n_{InP}))^3$. The cavity material is Indium Phosphide (InP) with a membrane thickness of 240 nm in a background of air. Topology-optimization was employed in the region marked by the red lines in Fig. 1(b) while shape-optimization was employed to tailor the shape of all inclusions by modifying their boundaries, as illustrated for a single inclusion using the orange circle in Fig. 1(b). Apart from the significant computational benefit, the employment of shape-optimization supports fabrication in that it is straight forward to limit complicated shapes and avoid extremely fine features, which are hard to fabricate accurately without significant effort dedicated to proximity effect correction.

The 2D device layout was designed using an iterative optimization procedure as described in Ref. [32]. An element-wise constant design field, $\xi(r)$, was defined for the topology-optimization region with one design variable per finite element used in the numerical discretization. For the shape-optimization region a set of 2D linear shape functions were used to parametrize the inclusions with the design variables being the associated basis-function coefficients. Standard filtering and thresholding [49] using continuation was employed for the topology-optimization domain, while a simple 1D diffusion filter [50] was employed for the shape-optimization variables individually for each inclusion to ensure a sufficiently slow variation of the curves describing the inclusions. In addition, connectivity and length-scale constraints [33, 51] were imposed on the topology-optimization domain to ensure fabricability. The physics was modelled using Maxwell's equations assuming time-harmonic fields and radiation boundary conditions. The design variables in the topology-optimization region were coupled to the physics model through a simple linear interpolation of the relative permittivity as, $\varepsilon(\xi(r)) = \varepsilon_{InP} + \xi(r)(\varepsilon_{Air} - \varepsilon_{InP})$, $\xi(r) \in [0,1]$. The model domain was excited using a Gaussian enveloped, plane-polarized incident electric field at the targeted wavelength ($\lambda = 1100$ nm). Under said excitation, the field intensity $|E|^2$ integrated over a central cylindrical target region with a radius of 15 nm in the device layer was maximized by iteratively changing the geometry. The target region was kept fixed as air as the goal of the design is to confine the electric field strongly inside the air region. The resulting device blueprint shown in Fig. 1(c) was optimized under fabrication constraints for a gap size of $g$ = 30 nm. All numerical simulations were carried out using a model implementation in COMSOL Multiphysics [52] on the DTU HPC system [53]. To reduce the computational complexity, the model domain was reduced by imposing and exploiting mirror symmetry along the two planes perpendicular to the device layer using perfect electric and magnetic conduction boundary conditions. A convergence study for the reference cavity was performed and results compared to a full model without symmetry conditions imposed to ensure numerical accuracy prior to both design and evaluation. The design process was carried out on a relatively coarse mesh using first order finite element basis functions to minimize computational costs in the design process. Meanwhile, all evaluations of the final design were carried out using a finer mesh and second order basis functions ensuring accuracy and showing no significant discrepancies between results obtained using the design model and the evaluation model.

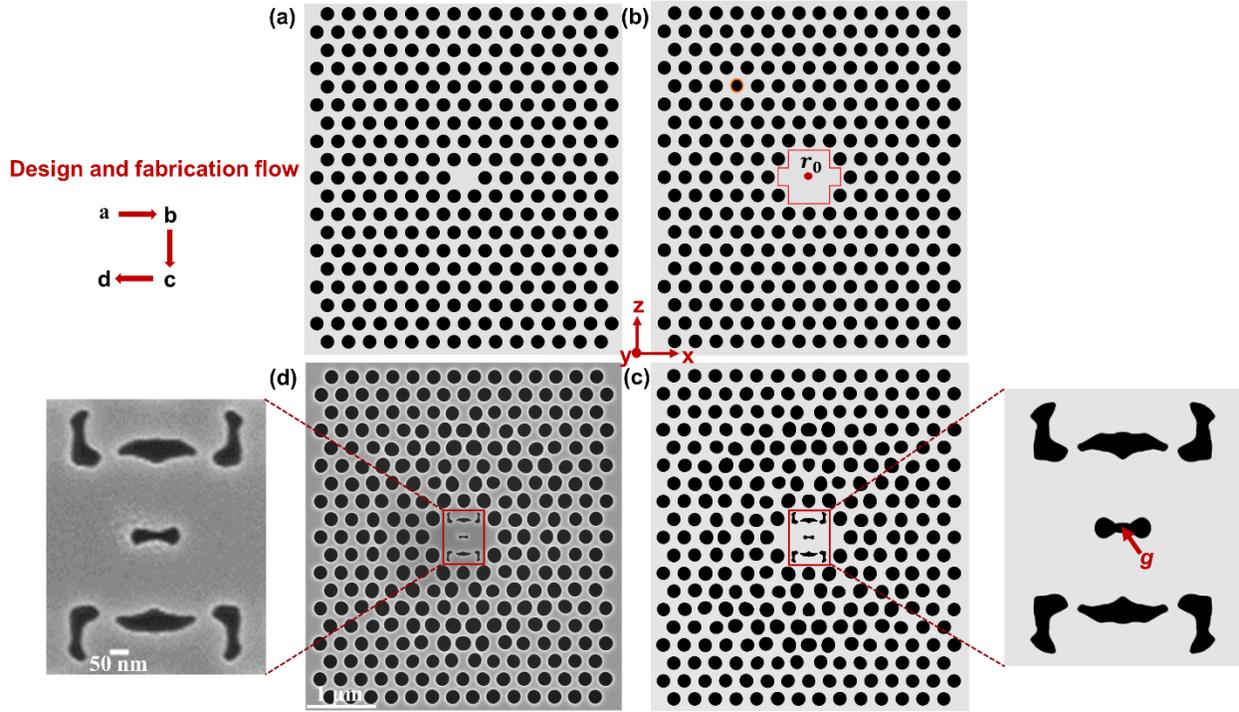

Fig.1 Design and fabrication process flow. (a) Reference PhC cavity supporting a mode with mode volume $V_{r_0} \approx 1.92\ (\lambda/(2n_{InP}))^3$. (b) The initial geometry used for the inverse design, using identical parameters to the reference cavity after removing the set of 6 holes surrounding the center defect. The sketch also includes a sketch of the designable parameters for both the shape- and topology-optimization (c) The final optimized cavity with gap size $g$ =30 nm and zoom-in for the central part. (d) SEM image of the fabricated cavity with $g \sim 28$ nm.

The samples were fabricated through the following steps: first, a hard mask of ~100 nm thick silicon nitride (SiNx) was deposited on the InP/SiO$_2$/Si substrate. Afterwards, 180 nm of e-beam resist CSAR 6200.09 was spin-coated onto the sample. The pattern was then defined through electron-beam exposure. To compensate for fabrication imperfections, each structure was repeated multiple times with varying degrees of shape correction. This process involved shrinking the exposed parts of the pattern by a specific length, known as the shrinkage constant (SC). The pattern was then transferred into the SiNx and InP layers by a two-step dry etching. Specifically, the etching of the nitride mask was performed with a plasma containing N2, SF6, CF4, and CH4, while InP was etched with a plasma containing HBr, Ar, and CH4 at an elevated temperature of 180°C. After the etching, the cavities were membranized by a buffered HF wet etching. This also removed the remaining nitride on the sample.

Figure 1(d) shows a scanning electron microscope (SEM) image of a fabricated structure with an estimated gap of $g \sim 28$ nm. The insert shows a zoomed view of the central part of the cavity, illustrating the agreement between the design blueprint and the fabricated device, which we observe for both the central region and the shape-optimized non-circular inclusions.

It is an interesting fact that inverse design aimed at increasing the optical field strength tends to lead to geometries with a single resonance at the frequency of interest. In this case, we also find that the electromagnetic response of the resulting geometry can be very well represented by a single quasinormal mode (QNM) over a wide bandwidth. The QNMs [54-57] – also known as resonant states [58-60] – are defined as solutions to the sourceless wave equation subject to suitable radiation conditions, such as the Silver-Müller radiation condition in this case. The radiation condition leads to complex frequencies of the form $\tilde{\omega} = \omega - i\gamma$, where $\gamma>0$ is the cavity field decay rate; the associated Q-value can be readily obtained as $Q = \omega/2\gamma$. Figure 2(a) shows the mode profile of the electric field-QNM of interest for the optimized design with $g$= 30 nm. The resonance wavelength is $\lambda \sim 1110$ nm, and the quality factor is Q~1245. It is evident that the field is strongly localized at the cavity center, where we find an effective mode volume [61] of

$V_{r_0} \approx 0.19 \, (\lambda/2n_{air})^3$. Figure 2(b) shows the single-QNM approximation to the Purcell factor in the center as a function of frequency along with an independent reference calculation. The precise agreement corroborates the validity of a single-QNM approximation.

## Characterization

We measured the resonance wavelengths and quality factors of the fabricated cavities using photoluminescence (PL) experiments. For the PL measurements, we used 637 nm laser as the source of excitation with the energy above the band gap of InP. A laser spot on the order of a few μm was achieved using an objective with NA=0.65, enabling excitation and signal collection from individual cavities. The signal was filtered using a long-pass filter and analyzed with a spectrometer equipped with a cooled InGaAs detector. All experiments used a polarizer to select z-axis polarization. Figure 2(c) shows the PL response from five fabricated clones of the same design with a gap of $g \sim 28$ nm. Circles in Fig. 2(c) show the experimental data while the solid lines are fits with Lorentzian functions from which we extracted the resonance wavelength $\lambda$, and quality factor Q = $\lambda$/FWHM, where FWHM is the full width at half maximum of the resonances. The average values are $\lambda_{av} = 1098.916$ nm and $Q_{av} = 388$, and the standard deviations, $\sigma_\lambda = 0.5$ nm and $\sigma_Q = 13$, confirm a robust and reproducible fabrication. A normalized experimental vs numerical polar plot of the far field emission at the target EDC mode is shown in Fig. 2(d) and shows that the mode is polarized along the z-axis as expected. The normalization in each case was performed by dividing the values by the peak value. We note that the observed cavity-enhanced PL emission is a result of the existence of a residual tail of below-bandgap InP emission, which extends into the cavity wavelength of ~ 1100 nm [62].

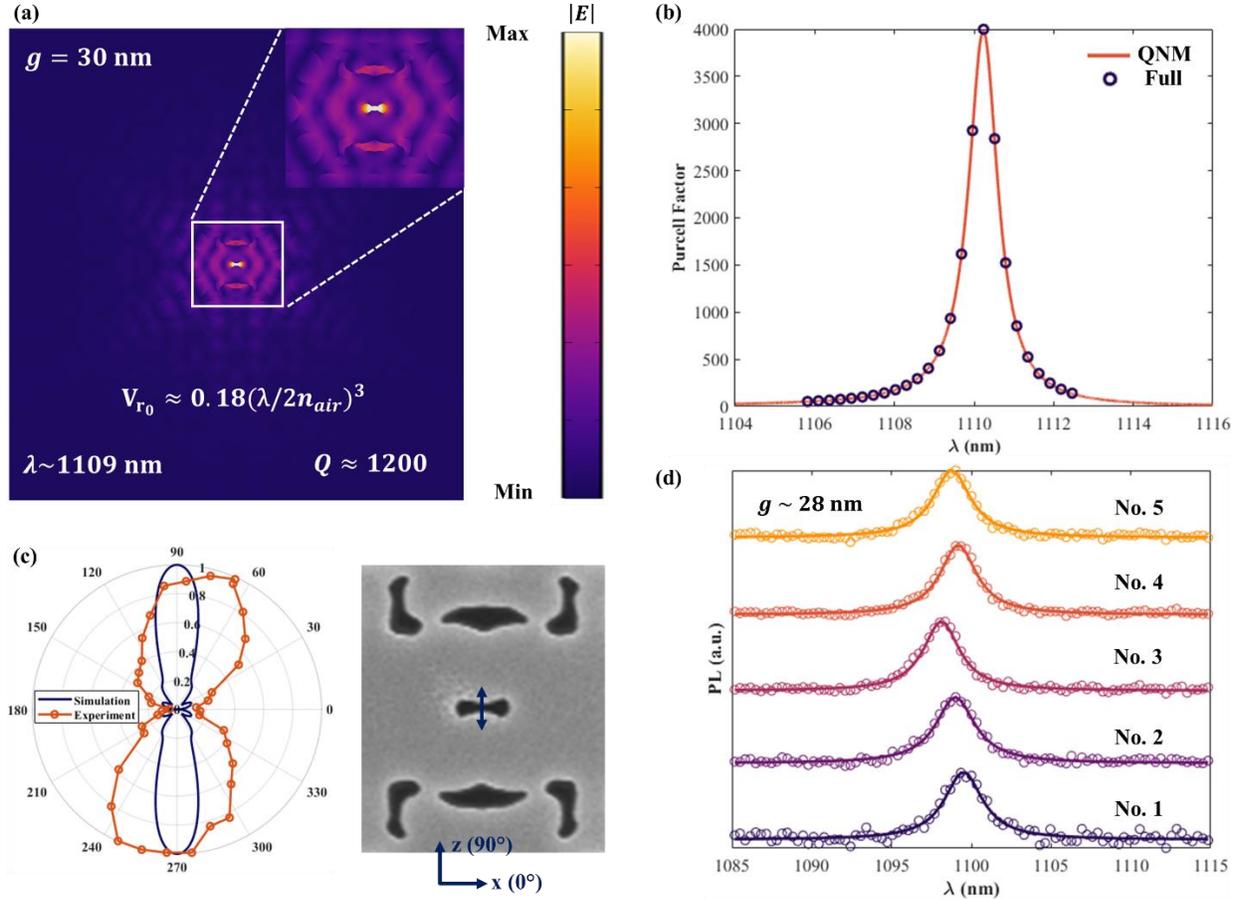

Fig.2. (a) Mode profile of the target EDC mode for the shape and topology-optimized cavity showing strong field at the tiny air gap of ~ 30 nm. (b) Spectral response of the structure. (c) PL spectra for five clones of nominally identical cavities. (d) Experimental vs. simulation polar plot (far field emission) at the target wavelength.

Next, we numerically explore the effect of uniformly shrinking all air-features of the design. In the upper row of Fig. 3, we show the central part of the changing geometry for four values of the gap size, $g$ =30 nm (original), 24 nm, 18 nm, and 6 nm. In the lower row, we show the associated mode profiles at each gap size. We observe a red-shift of the resonance wavelength with decreasing gap size, which is consistent with the expected shift from introducing additional high-index material in the cavity [63]. Surprisingly, the approach also leads to an improved spectral confinement of the field as evidenced by larger Q for smaller gap sizes. The mode volumes for the different values of $g$ are shown below the mode profiles. As expected [64], smaller gaps lead to stronger spatial confinement of the light and consequently smaller mode volumes. For the final gap size of $g$= 6 nm, the mode volume comes down to $V_{r_0} \approx$ 0.043 $(\lambda/2n_{air})^3$.

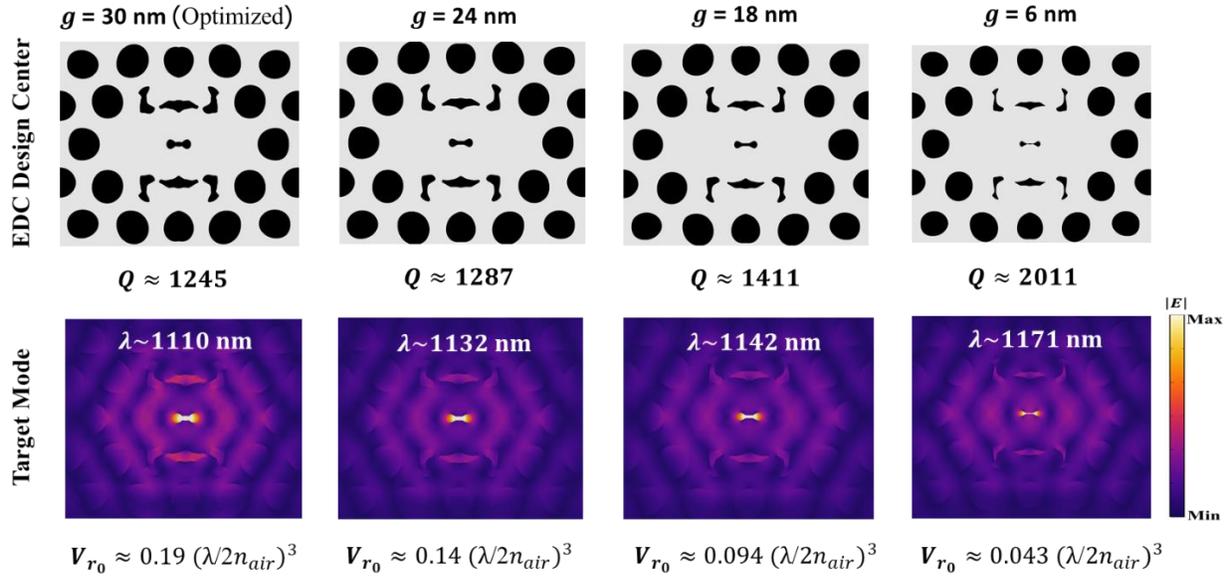

Fig.3. Top row: shrinking the gap size in the simulation. Bottom row: the associated computed mode profiles having the reported resonance frequencies, quality factors, and mode volumes.

Fig. 4(a) shows SEM images of the fabricated shape and topology-optimized cavities along with zoom-ins of the central feature. We studied gradually the effect of shrinking the gap from $g$~ 28 nm to $g$~ 6 nm and beyond, in which case we eventually closed the gap entirely. Fig. 4(b) shows the corresponding PL spectrum for cavities with $g$ ~ 28 nm, $g$ ~ 24 nm, $g$ ~ 22 nm, $g$ ~ 18 nm, $g$ ~ 13 nm, and two cavities with $g$ ~ 0 nm but having different holes size. The resonance wavelengths and quality factors of the cavities as a function of the gap size are summarized in Table. 1 below. The circles in Fig. 4(b) show the experimental data while the solid lines show fits using Lorentzian functions. Similar to the simulations, we observe a red-shift of the resonance with decreasing gap size. The trend in the experiments qualitatively agrees with the numerical calculations with a smaller shift observed in the experiments, which may be due to minor geometric discrepancies between blueprints and realized devices increasing with shrinkage. Regarding Q, a small improvement is observed when shrinking the gap size, and the experimental Q-value is about 3-times smaller than the numerical value at the same gap size, but the trend is similar to the numerical predictions of Q when shrinking the gap size. The data for $g$ ~ 6 nm is missing, because the sample was lost before the PL measurements.

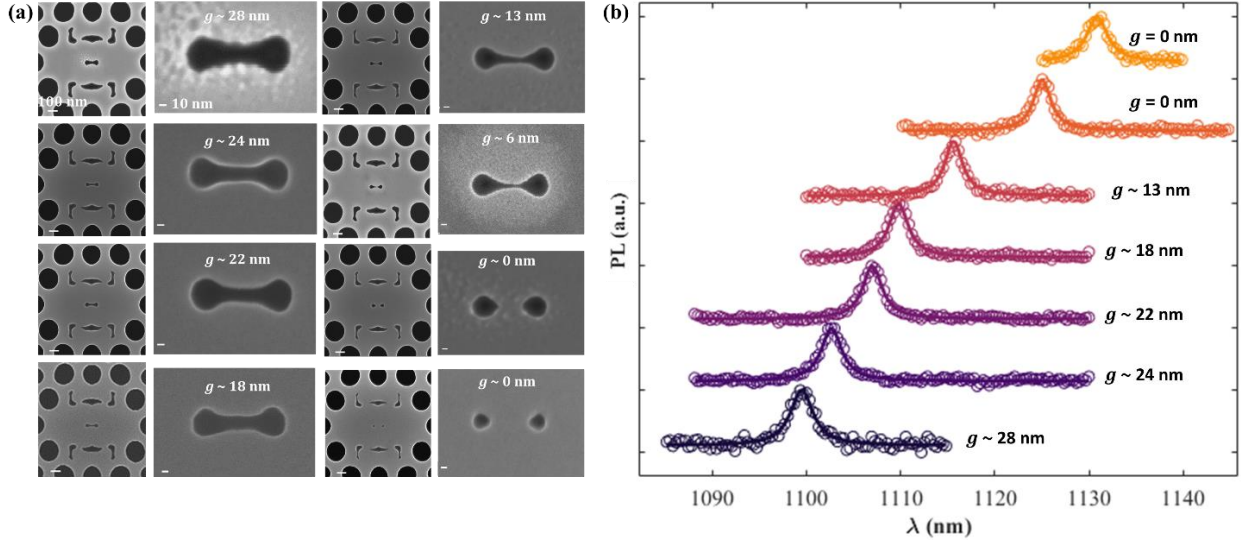

Fig.4. (a) SEM images of the shape and topology-optimized cavities after shrinking the gap. (b) Measured PL spectrum of five cavities with different gap sizes.

| $g\ (nm)$ | 28 | 24 | 22 | 18 | 13 | 0 | 0 |
|---|---|---|---|---|---|---|---|
| $\lambda\ (nm)$ | 1099.51 | 1102.67 | 1107.01 | 1109.79 | 1115.63 | 1125.03 | 1130.86 |
| Q | 381 | 419 | 446 | 448 | 488 | 501 | 420 |

Table.1. Resonance frequency $\lambda$ and quality factor Q as a function of the gap size for the shape and topology-optimized cavities with PL spectra shown in Fig. 4(b).

## Interfacing with colloidal QDs

Several methods for placing single QDs at well-defined positions in plasmonic or dielectric cavities [44-47] have been reported in the literature. In this last section, we report on our initial results from spin-coating colloidal Silver Sulfide ($Ag_2S$) QDs with diameters on the order of few nanometers on the fabricated cavities.

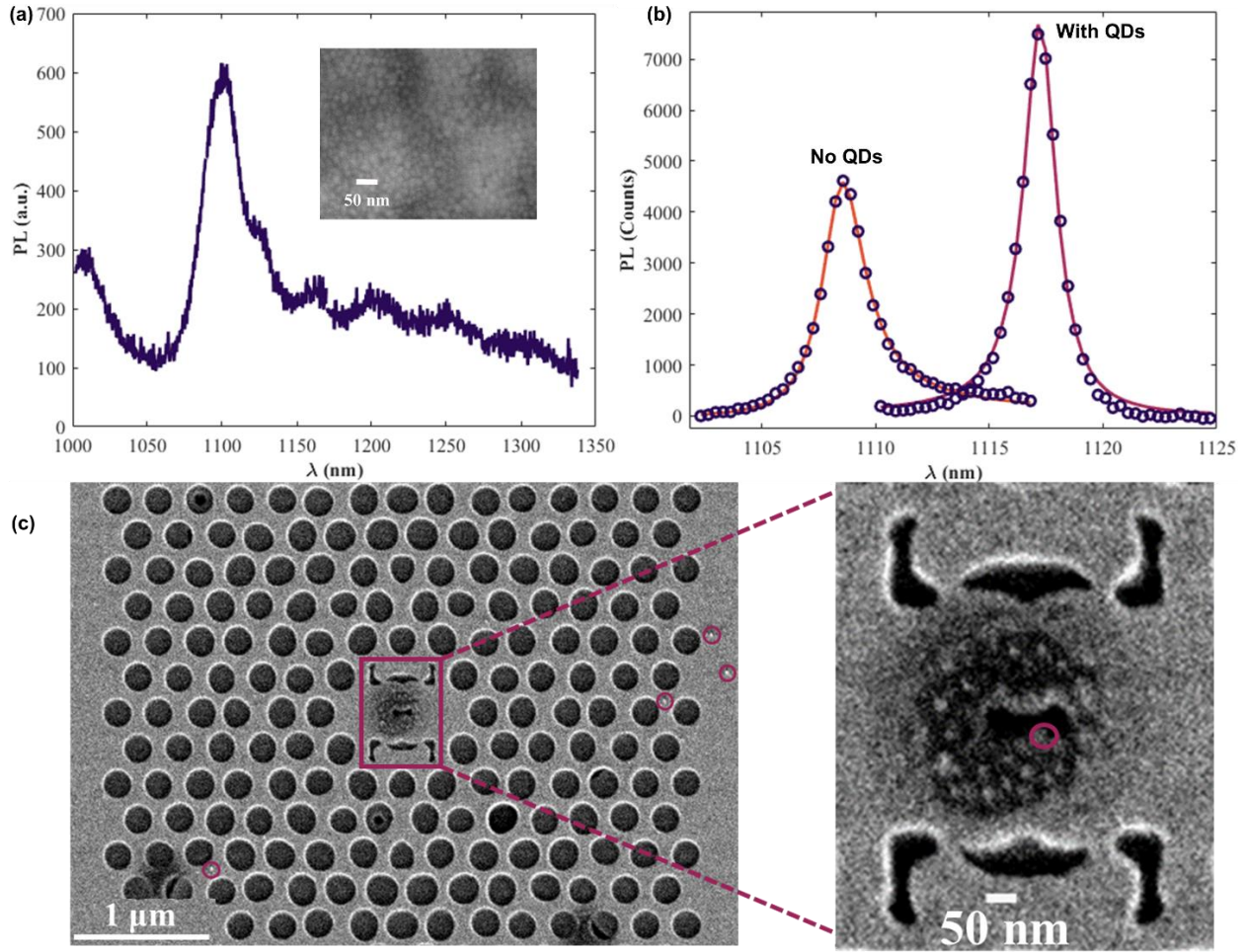

Fig.5. (a) PL spectrum for colloidal Ag2S QDs on Silicon substrate with the SEM image in the inset. (b) PL spectrum comparison of two similar cavities with and without QDs. (c) SEM image of a fabricated cavity with spin-coated colloidal Ag2S QDs, which are barely visible as small white elements in the figure. The zoom-in illustrates the QDs positioned in and around the gap in the center of the cavity.

Figure 5 (a) shows the PL spectrum for bare colloidal Ag2S QDs on a silicon substrate as shown in the SEM image in the inset. The QD PL spectrum has a dominant peak at ~ 1100 nm and shows a relatively large inhomogeneous broadening, which easily overlaps the target cavity mode wavelength. In the experiment, the QDs were spin-coated onto the sample by first placing a diluted droplet of QDs in solution on the sample and subsequently spinning in ~ 20 rpm for a few seconds. A SEM image of one of the cavities with spin-coated QDs is shown in Fig.5 (c). In this case, we were lucky to find multiple QDs around the central part of the cavity. Additionally, a few QDs were found far from the cavity center as marked with the red circles. Since these QDs are far from the cavity center, we expect them to not be significantly coupled to the resonant light, which is concentrated in the central gap, cf. Fig. 2a.

Figure 5b shows PL measurements for two nominally identical cavities with and without QDs. The measurements were taken immediately after each other with identical pump, and following our statistical analysis of the fabricated samples, we expect the spread in resonance frequency and Q-values to be much less than the observed differences, cf. Fig. 2c. The resonance wavelength is shifted from 1108.5 nm to 1117.3 nm. In addition, a narrowing in the spectral line width of the resonance is observed for the cavity with QDs with a reduction of the FWHM from 2.3 nm to 1.7 nm corresponding to an increase in the quality factor from Q=480 to Q=650. We attribute this shift to the presence of the QDs, but due to lack of reference measurements on different samples, we refrain from drawing strong conclusions about the nature of the coupling. Nevertheless, we note that the observed decrease in linewidth and consequent increase in count rate of the measured signal is consistent with a model in which the light-matter interaction of a weakly coupled QD-cavity system leads to an increase in lifetime of the cavity field [65-66]. For these measurements, we pump above the bandgap of InP and therefore expect most of the emission to be due to Purcell-enhanced recombination in the InP, even if the confinement of light in the gap region leads to a measurable effect on this recombination due to interaction with the QDs.

## Conclusions

We have presented our design and subsequent modeling, fabrication, and characterization of PhC slot cavities with deep sub-wavelength confinement of light with the aim of exploiting the resulting enhancement in light-matter interaction for interfacing with colloidal quantum dots. For the design, we employed a combined topology- and shape-optimization approach starting from a 2D PhC cavity as an initial guess to achieve enhanced electromagnetic field strength in a narrow air region at the cavity center while leaving sufficient space to enable positioning of quantum dots. The combination of shape and topology optimization enables computational tractability for the problem size while significantly simplifying the design. By constraining the minimum geometric length-scale of the design blueprint, we ensure high yield in fabrication with reliable resonance behavior of the cavities. Starting from the initial calculated values for the gap of 30 nm of $\lambda = 1113$ nm and Q~1257, numerical predictions show that both the wavelength and the quality factors are expected to increase when decreasing the gap size. In addition, a decrease in gap size enhances the relative field strength and lowers the effective mode volume for which we calculated values as low as V~ $0.09(\lambda/2n_{air})^3$ for the case of an 18nm gap. The significant shrinkage of the mode volume compensates for the moderate quality factor to achieve enhanced light-matter interaction in compact devices.

We successfully fabricated the optimized designs and characterized the samples to find a qualitative agreement with the expected changes in resonance wavelength and quality factor when shrinking the gaps. Moreover, we found a high reproducibility with standard deviations of the resonance frequency and the quality factor on the order of $\sigma_\lambda = 0.5$nm and $\sigma_Q$=13, respectively. In a last step, we performed spin-coating of colloidal Silver Sulfide QDs onto the sample and succeeded in finding a sample with QDs located around the central gap-region of the cavity. PL measurements on this sample show a shift and narrowing of the linewidth compared to that of a nominally identical reference sample, which is much larger than the shifts expected from the measured standard deviations. Hence, we attribute the measured effect to the interaction with the QDs, but due to lack of reference measurements on different samples we refrain from drawing any final conclusions on the nature of the interaction. Nevertheless, we consider the results to be a proof of principle showing the viability of interfacing colloidal QDs with the optimized PhC cavities. With future optimization of the spin-coating process it may be possible to achieve consistent positioning of single QDs in the cavity center and thereby exploit the material platform for practical applications such as the development of single photon sources.

## ASSOCIATED CONTENT

## AUTHOR INFORMATION

### Corresponding Author

**Mohammad Abutoama** − *DTU Electro, Technical University of Denmark, Ørsteds Plads, building 343, 2800 Kgs. Lyngby, Denmark; NanoPhoton - Center for Nanonphotonics, Ørsteds Plads, building 345A, 2800 Kgs. Lyngby, Denmark; orcid.org/0000-0002-4286-0434; Email: moabo@fotonik.dtu.dk*

**Rasmus Ellebæk Christiansen** − *Department of Civil and Mechanical Engineering, Technical University of Denmark, Nils Koppels Allé building 404, 2800 Kgs. Lyngby, Denmark; NanoPhoton - Center for Nanonphotonics, Ørsteds Plads, building 345A, 2800 Kgs. Lyngby, Denmark; orcid.org/0000-0003-4969-9062; Email: raelch@dtu.dk*

**Adrian Holm Dubré** − *DTU Electro, Technical University of Denmark, Ørsteds Plads, building 343, 2800 Kgs. Lyngby, Denmark; NanoPhoton - Center for Nanonphotonics, Ørsteds Plads, building 345A, 2800 Kgs. Lyngby, Denmark; Email: adrianholmdubre@hotmail.dk*

**Meng Xiong** − *DTU Electro, Technical University of Denmark, Ørsteds Plads, building 343, 2800 Kgs. Lyngby, Denmark; NanoPhoton - Center for Nanonphotonics, Ørsteds Plads, building 345A, 2800 Kgs. Lyngby, Denmark; Email: menxi@dtu.dk*

**Jesper Mørk** − *DTU Electro, Technical University of Denmark, Ørsteds Plads, building 343, 2800 Kgs. Lyngby, Denmark; NanoPhoton - Center for Nanonphotonics, Ørsteds Plads, building 345A, 2800 Kgs. Lyngby, Denmark; Email: jesm@dtu.dk*

**Philip Trøst Kristensen** − *DTU Electro, Technical University of Denmark, Ørsteds Plads, building 343, 2800 Kgs. Lyngby, Denmark; NanoPhoton - Center for Nanonphotonics, Ørsteds Plads, building 345A, 2800 Kgs. Lyngby, Denmark; orcid.org/0000-0001-5804-1989; Email: ptkr@dtu.dk*

**Funding/ ACKNOWLEDGMENT**

This work was supported by the Danmark Grundforskningsfond through the research center "NanoPhoton", grant number DNRF147. M.A. was partially supported by the Planning and budgeting committee of the Israeli council for higher education (VATAT). We are grateful to Frederik Schröder and Yuri Berdnikov for helpful discussions and guidance.

**Notes**

The authors declare no competing financial interest.